\documentclass[conference,10]{IEEEtran}
\IEEEoverridecommandlockouts

\usepackage{url}
\usepackage{amsmath}
\usepackage{amsfonts}
\usepackage[ruled,longend,boxruled]{algorithm2e}
\usepackage{graphicx}
\usepackage{textcomp}
\usepackage{xcolor}
\def\BibTeX{{\rm B\kern-.05em{\sc i\kern-.025em b}\kern-.08em
    T\kern-.1667em\lower.7ex\hbox{E}\kern-.125emX}}

\begin{document}

\title{Performance Models for a Two-tiered Storage System}

\author{\IEEEauthorblockN{Aparna Sasidharan}
\IEEEauthorblockA{\textit{Computer Science Dept} \\
\textit{IIT}\\
Chicago, USA}
\and
\IEEEauthorblockN{Xian-He}
\IEEEauthorblockA{\textit{Computer Science Dept} \\
\textit{IIT}\\
Chicago, USA}
\and
\IEEEauthorblockN{Jay Lofstead}
\IEEEauthorblockA{\textit{Computer Science} \\
\textit{Sandia National Lab}\\
New Mexico, USA}
\and
\IEEEauthorblockN{Scott Klasky}
\IEEEauthorblockA{\textit{Computer Science} \\
\textit{Oak Ridge National Lab}\\
Tennesse, USA}
}

\maketitle

\begin{abstract}
This work describes the design, implementation and performance analysis of a distributed two-tiered storage software. The first tier functions as a distributed software cache implemented using solid-state devices~(NVMes) and the second tier consists of multiple hard disks~(HDDs). We describe an online learning algorithm that manages data movement between the tiers. The software is hybrid, i.e. both distributed and multi-threaded. The end-to-end performance model of the two-tier system was developed using queuing networks and behavioral models of storage devices. We identified significant parameters that affect the performance of storage devices and created behavioral models for each device. The performance of the software was evaluated on a many-core cluster using non-trivial read/write workloads. The paper provides examples to illustrate the use of these models.
\end{abstract}

\section{Motivation\label{motiv}}
Scientific computing applications~\cite{nbody},~\cite{md} have shown good scalability on High-Performance Computing~(HPC) machines with many-core processors. These machines designed using heterogeneous nodes~(CPUs and GPUs) are capable of several teraflops/sec or exaflops/sec throughput for computations. Their performance drops with IO intensive workloads such as streaming~\cite{stream} and archived data applications.
In this article, we discuss a system that is suitable for processing archived data. Parallel IO performance on clusters depends on factors such as layout of files on storage devices~(number of devices, file sizes, file partition sizes), read/write bandwidth and load distribution across storage devices, communication network and number of processes~(readers/writers). It is difficult for storage systems to match the compute throughput of new machines which generate high volumes of IO requests at fast rates due to their increased concurrency. Storage system designs with shared HDDs are not sufficient to match these new request rates in spite of MPI-IO optimizations that reduce the number of IO requests. Parallel IO on such systems is affected by data transfer costs over the network and load from other users~\cite{mpiurl}. Tiered storage designs which cache frequently used data on faster solid-state devices~(tier 1) and use HDDs for permanent storage~(tier 2) have higher throughput~\cite{randomIO}. HPC machines use the tiered approach due to economic reasons and for fault-tolerance. Distributed caching and tiering storage systems have been areas of active research~\cite{dahlin},~\cite{gms},~\cite{hpccache},~\cite{staging}. Caching remote files on local clients differs from caching files on solid-state devices in clusters. File transfers from remote servers would take seconds, while IO accesses~(KB/MB) from disks in a cluster can be performed in milliseconds. We address the second problem here that is restricted to HPC clusters. There are several unknowns in using tiered storage such as request distribution, cache size, cache line size, number of processes and data eviction policies. 

In this work we describe performance models that are useful in analyzing and configuring a two-tiered storage system based on application requirements. The two storage devices differ in speeds due to mechanical differences, concurrency and read/write bandwidth gaps. Our models help to determine the best way to compose these two device types to match the overall requirements of an application. We modeled the flow control of the system using quantities such as arrival rates, response times, waiting times and service times. The two devices were modeled by defining parameters for load distribution, synchronization, concurrency and communication costs. Most of the recent available performance data for tiered HPC storage systems are post data staging~\cite{staging}. In this paper, we used tier 1 as an \emph{inclusive} cache and forwarded misses to tier 2. The measured performance includes the cost of page misses. We used NVMes~\cite{nvperfr} as tier 1 solid-state devices and the HDF5 parallel IO library~\cite{hdf5} for tier 2 file accesses from HDDs. We developed an Online Learning~(OL) algorithm for cache replacement. This algorithm learned the popularity and temporal locality of the IO traffic and minimized total number of IO requests. Our hypothesis is that a mix of cache replacement algorithms will perform better for complex IO traffic. We considered two categories of IO request prefetchers in our implementation : stream identifiers and Markov chains~\cite{markovts}.
Related work is described in detail in section~\ref{3}, the software architecture and OL page replacement in section~\ref{4}. A short discussion of machine architectures is provided in section~\ref{8} and the performance models are discussed in section~\ref{5}. Experiments and conclusions are described in sections~\ref{6} and \ref{7}. The contributions of this paper are listed below : 
\begin{itemize}
\item An end-to-end performance model for two-tiered storage systems based on queuing networks.
\item Behavioral models of two types of storage devices based on their load distribution and parallelism.
\item Implementation of OL algorithm for cache replacement in tiered storage and its performance evaluation using different types of traffic models.
\item Performance evaluation of the full software benchmark on a recent cluster.
\end{itemize}	
The results of experiments, observations and conclusions drawn from them will be useful for the HPC community.

\section{Related Work~\label{3}}
Previous work on data movement in multi-tiered storage architectures can be found in~\cite{storagehier}. They haven't addressed approaches for minimizing data movement between tiers. Other designs for tiered storage such as those described in~\cite{tiered2} and~\cite{tiered3} have used novel replacement policies. A general approach for such designs is to use priority queues for storing data in the tiers. Data movement between tiers depends on the IO traffic and the priority function. There were efforts within the scientific community to model common page access patterns, referred to as traffic models~\cite{zipf},~\cite{iopatterns}. Majority of this work was meant for sequential storage systems in which multiple clients use a single interface to access data. Caching or tiering in distributed systems with independent caches was first described by Dahlin et al.~\cite{dahlin},~\cite{dahlinps},~\cite{hpccache}. Global page replacement algorithms are described in~\cite{gms},~\cite{sumcache}. An evaluation of remote caching algorithms for different workload types is discussed by Leff et al.~\cite{leff}. We did not consider global replacement algorithms in this work.
Recent work in multi-tiered storage includes the use of solid-state devices as distributed cache~\cite{d3n},~\cite{orthus}. The authors of~\cite{orthus} and \cite{reflex} have developed performance models for NVMes. But their models do not fully address the overheads from accessing random file offsets.~\cite{Hermes} has also used NVMes as distributed cache, but their design has a centralized metadata manager which creates overheads. Solid-state devices are being widely used as fast storage in HPC clusters, such as Proactive Data Containers~\cite{PDC}, DAOS~\cite{daos}. They are commonly used for buffering~\cite{staging} and rarely as demand-paged caches.

Data movement or page replacement algorithms have been studied in depth~\cite{henessay},~\cite{belady},~\cite{tarjan},~\cite{pagerank},~\cite{pagemig},~\cite{page1},~\cite{acme}. These papers include both theoretical and experimental results. Besides LRU and LFU, some of the widely discussed page replacement algorithms are LRU-K~\cite{lruk},random, MIN,FIFO~\cite{belady} and WS~\cite{WSP}. The optimal page replacement algorithm is MIN, described in~\cite{belady}. That LRU is as good as optimal for slow evolving sequences was proved by~\cite{pagelru},~\cite{belady}. Eviction algorithms based on scoring functions were discussed by~\cite{score}. For any access sequence, an ideal page replacement algorithm is one that has full knowledge of the sequence. A page replacement algorithm can peek into the short-term future for evicting pages that are least likely to be used in the near future~\cite{belady}. For simple strided sequences, near future page accesses can be predicted using stream identifiers or prefetchers~\cite{henessay}. Online learning algorithms for page replacement were explored by~\cite{mlcache},~\cite{mlcache1}. The implementation described by~\cite{mlcache} is for two experts only. \cite{mlcache1} uses online convex optimization, which is more flexible. But they have not addressed the computation of time-varying popularity and utility functions.
Complex page access sequences can be modeled using Markov chains~\cite{markovp}. The most probable next states are predicted using the current state and transition probabilities. Markov chains are better at recognizing non-trivial sequences than stream identifiers that compute differences between address offsets. \cite{mlpatterns} used unsupervised learning to predict page accesses from complex Markov chains. 

IO performance modeling is a challenging problem. Analytical models of storage systems were created to understand quantities such as rotational delay, seek time and IO bandwidths for different IO sequences, request sizes and hardware. Poisson processes were used to model IO requests and queuing theory to analyze average waiting time and service rate~\cite{queuingsystem}. But with complex IO systems used in shared environments, these models are inadequate. Service times can show significant variation with load on shared disks. Recent work on storage systems has used supervised learning to model quantities such as IO read/write times and server utilization~\cite{mlhdd1},~\cite{mllb2}. Parallel IO libraries such as HDF5 have used supervised learning models for predicting better file layouts for applications~\cite{autotuning}. Performance models for NVMes are recent~\cite{orthus},~\cite{nvmepf}. These models have captured some features of NVMe performance. \cite{randomIO},~\cite{reflex} and \cite{libra} have developed performance models for IO resource management in multi-tenant environments. But the models, metrics and QOS requirements are not fully developed.
\section{Software Architecture\label{4}}
This section describes the software architecture of our IO benchmark that was used for the experiments in section~\ref{6}. We used MPI for spawning processes and for collective communication. Every MPI process has a corresponding posix file functioning as tier 1 cache located in its nearest NVMe. Each posix file is divided into $N$ $m$-byte cache lines~(pages); the values of $N$ and $m$ being configurable. The posix files are accessed by mapping their pages to CPU memory. Processes transferred data to their NVMe files using PCIe network. Each cache line has \emph{index}, \emph{tag}, \emph{valid} and \emph{dirty} bits. Our implementation is a demand-driven, fully-associative, write-back cache and transitions between cache states are similar to those found in CPUs~\cite{henessay}. The cache lines are multi-reader, single-writer and mutual exclusion is guaranteed by using read-write locks on the states. To reduce lookup time, cache states are stored in CPU memories and data on NVMe files. Tier 1 can also be described as a single copy inclusive distributed cache because we do not allow data replication. Replication would have required the implementation of a cache-coherency protocol and metadata management. Page migration between caches is also not allowed in our current implementation.

Pages are distributed across processes using a suitable mapping policy. We implemented widely used mapping policies such as round-robin, random, block and block-cyclic~\cite{cachemapping}. A combination of file number and $page\_no= \frac{FILE\_OFFSET}{CACHE\_LINE\_SIZE}$ were used as \emph{index} and \emph{tag} fields. The random policy maps page numbers to processes using hash functions. The definitions of other mapping policies are consistent with their descriptions in literature. A suitable mapping function may be chosen based on the correlation between page accesses in a distributed application. For example, if all processes share a common set of pages, and if requests are uniformly distributed, random mapping will provide good load balance. Block mapping will minimize inter process communication for exclusively accessed pages.
\begin{figure}[!tbph]
\begin{center}
\includegraphics[width=2.4in,height=2.2in]{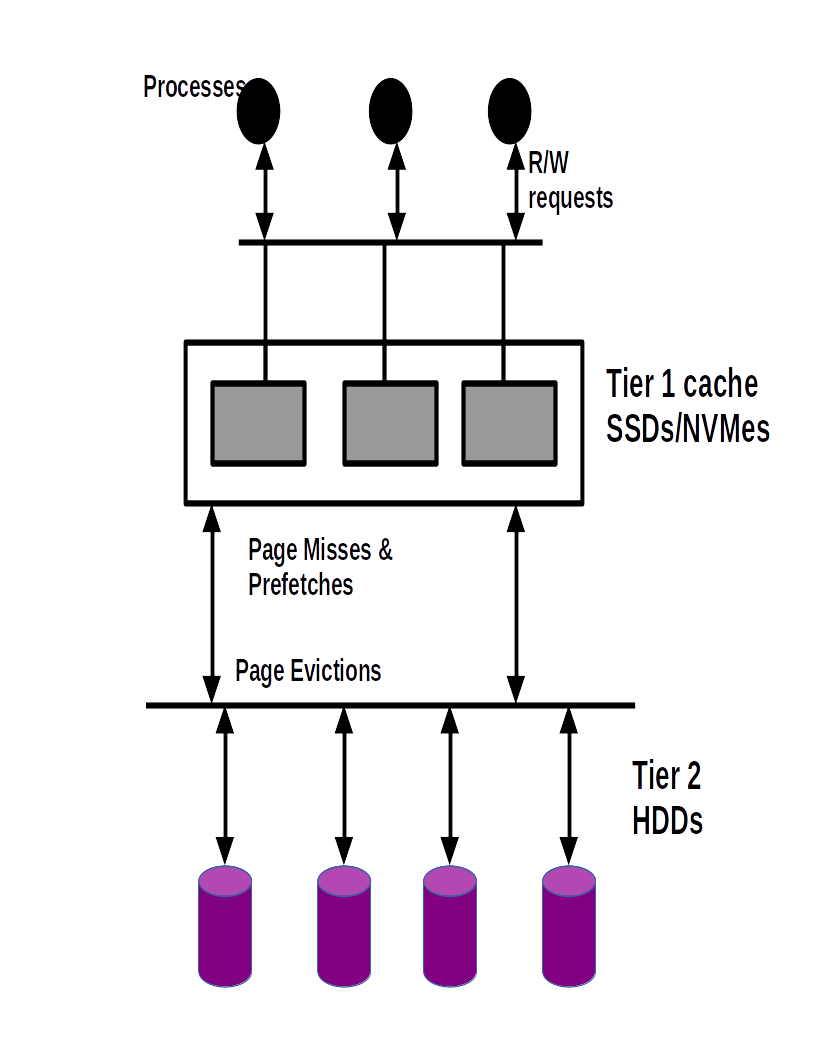}
\caption{Two tier Storage : Software Architecture\label{fig:fig1}}
\end{center}
\end{figure}

\begin{figure}[!tbph]
\begin{center}	
\includegraphics[width=3in,height=2.4in]{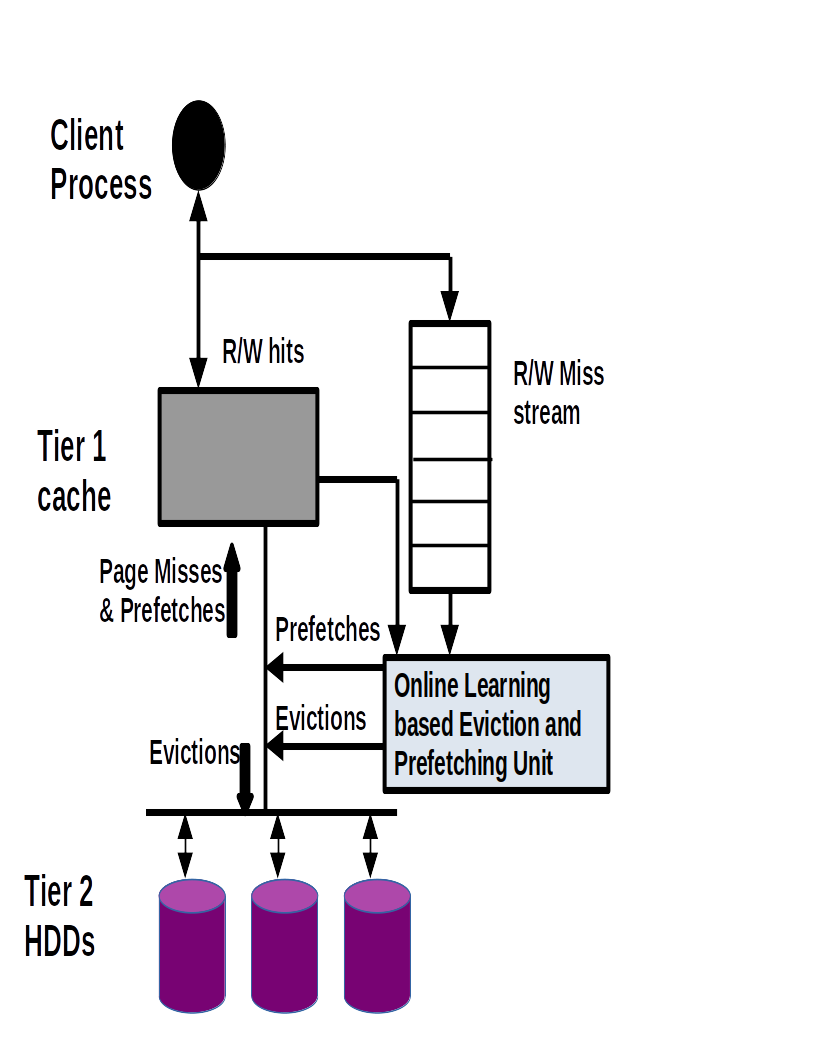}
\caption{Two-tier Storage : Software Architecture and Data Movement\label{fig:fig2}}
\end{center}
\end{figure}

The tier 2 implementation consists of an interface for submitting IO requests generated by tier 1 page misses. MPI threads generate read/write requests of $s$ bytes each. In our implementation $t+s \leq CACHE\_LINE\_SIZE~(page size)$, where $t$ is the base address of an IO request. These requests are forwarded to the distributed cache via multi-threaded RPCs provided by Mercury~\cite{mercury}. Every process has a dedicated IO thread and an IO queue. Page misses in tier 1 are forwarded to IO queues which are shared between client and IO threads. IO threads poll the IO queues, and distribute page requests across processes according to the mapping functions described earlier in this section. Page misses are serviced by IO threads and pages are placed in tier 1 slots. The OL~\cite{onlinelearning} cache replacement algorithm is activated when caches are full. The software architecture is described in figures~\ref{fig:fig1} and \ref{fig:fig2}. Figure~\ref{fig:fig1} shows the tier 1 distributed cache and the tier 2 HDDs, and figure~\ref{fig:fig2} shows the components involved in the movement of data between tiers for a single process. Eviction decisions depend on the observed page miss streams and the most recent cache states. Every cache line has frequency counter and timestamp fields associated with its state. These fields are updated during cache accesses. In our design, page misses are prioritized over prefetches. The most recent miss stream is used to generate prefetch requests using a stream identifier. Prefetched pages are stored in separate prefetch buffers and follow the same mapping function as the cache. On a miss, a page is first located in the prefetch buffer. If found, it is removed and promoted to the cache. In our current implementation, prefetching is performed only if there are empty slots in the prefetch buffer. The width of the buffer decides the maximum number of prefetches in any iteration. 
In our current implementation, cache sizes are fixed. Resizing was useful in implementations where remote caches were located in CPU RAMs~\cite{talus}. It will be relevant in multi-tenant implementations which cache several files.     
\subsection{Online Learning for Data Movement}
\begin{figure}[!tbph]
\begin{center}
\includegraphics[width=2.2in,height=1.8in]{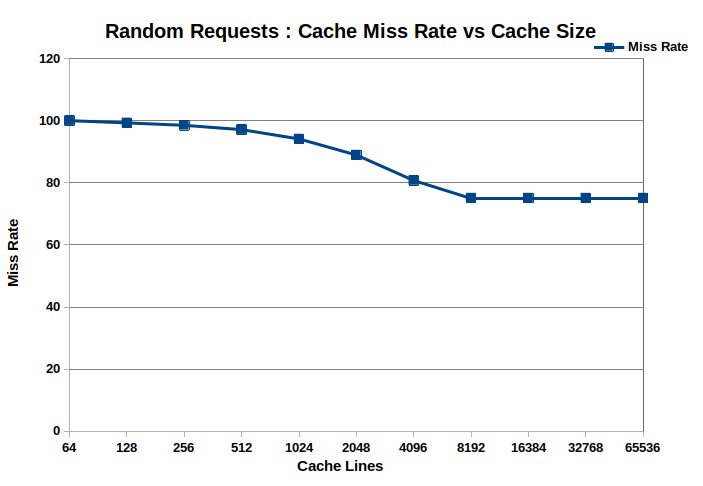}
\caption{Capacity Misses : Miss Rate vs Cache Size for 1 MPI process\label{fig:fig3}}
\end{center}
\end{figure}

Assuming all page requests are the same size, data movement can be reduced by minimizing~\cite{fileprefetching} page misses.
Evictions and prefetches modify the page miss stream sequence. A good page eviction algorithm and a prefetcher that accurately identifies data access patterns will minimize the total number of page misses. If prefetching is performed when IO threads are idling~(no page misses), it will reduce the total waiting time for IO requests. Since caching is similar to promote/demote operations in splay trees, the effectiveness of eviction algorithms and prefetching can be studied using amortized analysis. Figure~\ref{fig:fig3} shows the plot of cache miss rate with increasing cache size for 1 MPI process using a random read workload. These experiments were performed on the Delta Supercomputer~\cite{delta} using our IO benchmark. The cache miss rate function matches expected behavior observed in CPU caches and other cooperative cache implementations.   

In our IO benchmark, we divided the polling iterations of IO threads into epochs and set epoch width equal to $4$ iterations. We implemented $3$ cache replacement policies : Least Recently Used~(LRU), Least Frequently Used~(LFU) and Random and used them as experts in a weight-sharing OL algorithm~\cite{onlinelearning},~\cite{ws}. 
If a page miss was generated for a page that was previously evicted in the same epoch, it is considered as a \emph{misprediction} and the algorithm is penalized for its decisions. Penalties are computed and the weights associated with the eviction experts are adjusted. These weights are converted to probabilities and the expert with the highest probability is chosen. Since the OL algorithm described here is as fast as the experts, we chose low-overhead policies as experts. Their decisions can be computed by reading the current cache state. We used timestamps for LRU and frequency counters for LFU. Like discussed in the previous sections, all experts are local replacement algorithms without the need for global consensus.  

\begin{algorithm}
\small
\DontPrintSemicolon
\KwIn{PageMisses $m$}
\KwIn{iter $t$}
\KwOut{Victim}
\SetKwBlock{Begin}{procedure}{end procedure}
\SetKwFunction{CopyArray}{CopyArray}
\SetKwFunction{InitArray}{InitArray}
\SetKwFunction{NumExperts}{NumExperts}
\SetKwFunction{ChooseExpert}{ChooseExpert}
\SetKwFunction{EvictExpert}{EvictExpert}
\SetKwFunction{AddDecision}{AddDecision}
\Begin($\text{GetVictim} {(} n {)}$)
{
 $n \gets \NumExperts{}$; 
 $p \gets \ChooseExpert{}$\tcc{Highest Probability}
 $\InitArray{evicts,n}$\;

\For {i $\in$ n}
{
$v \gets \EvictExpert{i}$;$evicts[i] \gets v$ \;
$\AddDecision{predictions[i],v}$
}
\Return {evicts[p]}\;
\caption{GetVictim \label{alg:alg2}}
}
\end{algorithm}

\begin{algorithm}
\small
\DontPrintSemicolon
\KwIn{PageMisses $m$}
\KwIn{iter $t$}
\KwOut{void}
\SetKwBlock{Begin}{procedure}{end procedure}
\SetKwFunction{CopyArray}{CopyArray}
\SetKwFunction{InitArray}{InitArray}
\SetKwFunction{NumExperts}{NumExperts}
\SetKwFunction{GetProb}{GetProb}
\SetKwFunction{GetRecDec}{GetRecDec}
\SetKwFunction{GetEpochWidth}{GetEpochWidth}
\Begin($\text{WeightAdjust} {(} n {)}$)
{
$n\gets\NumExperts{}$\;
\InitArray{prevwts,n} \tcc{Initialize,Copy}
\CopyArray{prevwts,weights,n} \;
pred $\gets$ \GetRecDec{}\tcc{Decisions(epoch)}
\InitArray{mispred,n} \;
prob $\gets$ \GetProb{} \tcc{Probability vector}

\For {p $\in$ m}
{
\For {i $\in$ n}
{
\If {p $\in$ pred[i]}
{
  $mispred[i] \gets mispred[i]+1$ \;
}
}
}
\tcc{Exit if epoch not expired}
\If {iter $==$ 0 | iter $\mod$ \GetEpochWidth{} $\neq$ 0} 
{
   \Return;
}

\tcc{Adjust weights \& probabilities}
\For {i $\in$ n}
{
  $l \gets mispred[i]$\;
  $d \gets \beta^l$\;
  $weights[i] = weights[i]-weights[i]*d$\;	
}

$s \gets 0$ ;\

\For {i $\in$ n}
{
  $s \gets s+prevwts[i]-weights[i]$ ;\
}

$s \gets \frac{s}{n}$ ;\
$den \gets 0$;\
\For {i $\in$ n}
{
$weights[i] \gets weights[i]+\alpha*s$;\
$den \gets den + weights[i]$;\
}

\For {i $\in$ n}
{
 $prob[i] \gets \frac{weights[i]}{den}$ ;\
}

\caption{WeightSharing : Weight Adjust\label{alg:alg1}}
}
\end{algorithm}

The weight-sharing algorithm used for evictions is described in algorithms~\ref{alg:alg1} and \ref{alg:alg2}. Algorithm~\ref{alg:alg1} describes the weight adjustment function that is based on counting the number of mispredictions made by each expert. The predictions made by experts are stored in a prediction vector~(algorithm~\ref{alg:alg2}). The prediction vector is cleared every $EPOCH\_WIDTH$ iterations to avoid mixing history from distant past. If the number of mispredictions is less than $THRESHOLD*miss\_count$, then they are ignored. $miss\_count$ is the number of page misses in an epoch. In our experiments, we set $THRESHOLD=0.25$. 
The algorithm chooses between experts according to their probabilities. Weights are adjusted in intervals of duration $EPOCH\_WIDTH$,to avoid swift changes between experts. We found page replacement to be a problem suitable for OL with experts. OL was previously used for cache replacement by ~\cite{mlcache1},~\cite{acme}. But they did not show the benefits of learning using multiple experts with different traffic models.  

\section{System Architecture~\label{8}}

\begin{figure}[!tbph]
\includegraphics[width=2.6in,height=1.8in]{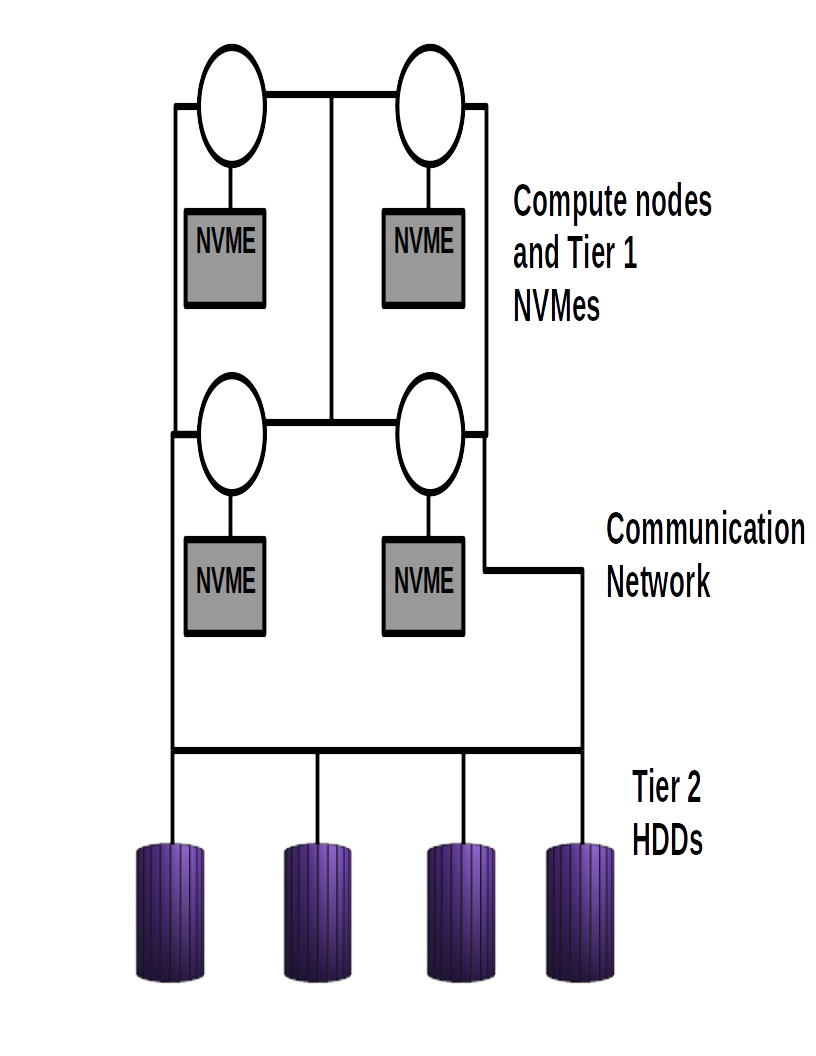}
\caption{Example of a cluster with two-tiered storage~\label{fig:fig4}}
\end{figure}

Increasingly production HPC clusters are adding tiers to their storage systems~\cite{d3n} for improving IO performance. Some of these machines have NVMes attached to every compute node along with slower HDDs and tapes. Compute nodes may use PCI or RDMA to communicate with their local NVMes. This may not be an economical design for large clusters and will lead to low utilization of NVMes for most compute intensive HPC applications. Some large HPC clusters have special nodes with attached NVMes to aggregate IO requests. Remote aggregation of short requests on a network of NVMes is an economical design for two-tiered storage. Sharing of NVMes by multiple HPC applications improves storage utilization. But this design has to pay the overhead of transferring large volumes of IO requests over the network for IO workloads. In this paper, we have restricted ourselves to medium-sized clusters in which every compute node has an attached NVMe, shown in the diagram in figure~\ref{fig:fig4}. These NVMes across compute nodes constitute the tier 1 distributed cache. Processes can access data stored on remote NVMes using the high-speed communication network. The example in figure~\ref{fig:fig4} has used HDDs for tier 2 and they are shared among all the nodes of the cluster. The data aggregated in NVMes are transferred to HDDs as large messages over the network. For large machines with several users and running high throughput concurrent applications, inter process communication can become an issue even with tiered storage. Although the discussion in this paper has used HDDs in tier 2, they may be substituted by another tier of NVMes. The design of a storage architecture for a machine is usually made after considering several factors including machine size, types of targeted applications, communication network, topology, processor architecture and budget~\cite{d3n},~\cite{costmodel}.

\section{Performance Models~\label{5}}
\begin{figure}[!tbph]
\begin{center}	
\includegraphics[width=2.6in,height=2.2in]{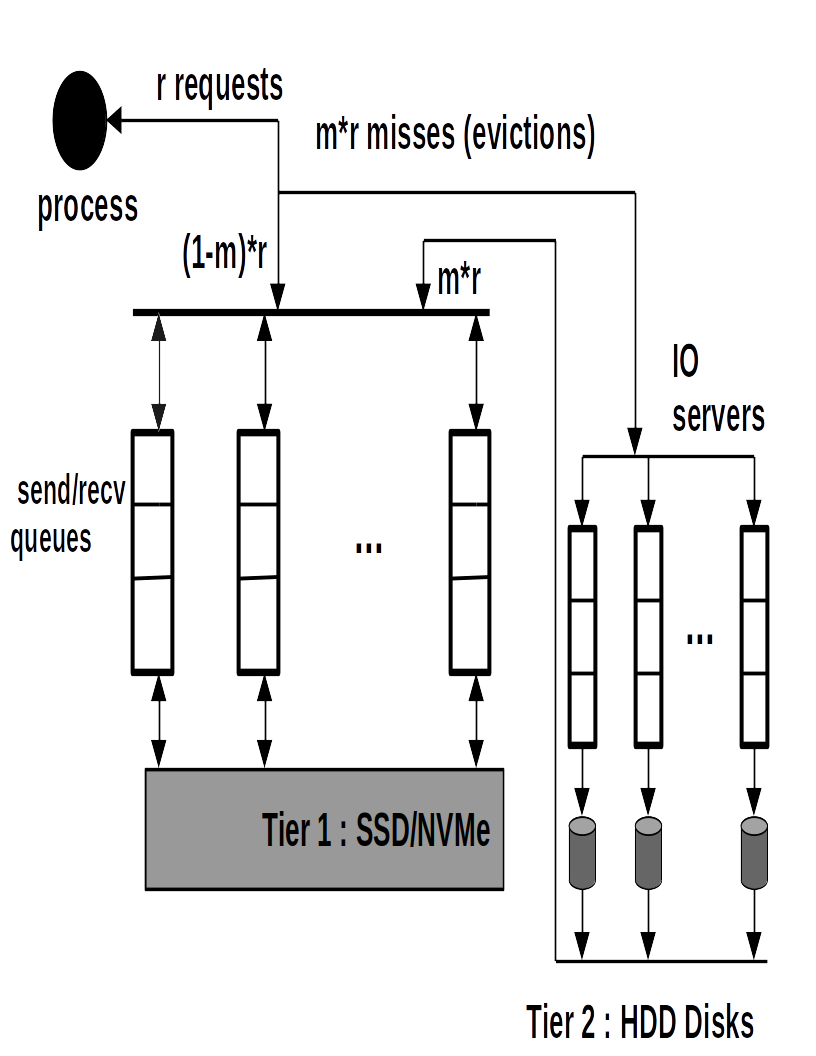}
\caption{Two-Tiered Storage : Queuing Network per process~\label{fig:fig5}}
\end{center}
\end{figure}
The two tiers can be modeled using a network of queues~\cite{kelly}~\cite{queuingtheory}, shown in figure~\ref{fig:fig5} where the inputs to tier 1 queues are read/write requests. Tier 1 hits exit the system and misses enter tier 2 queues. They are serviced by tier 2 and re-enter tier 1. Additional requests generated by tier 1 such as evictions and prefetches are serviced by tier 2. Pages evicted from tier 1 are serviced by tier 2 and exit the system.
The arrival of requests to tier 1 can be modeled using a random variable, with expected arrival rate $E(\lambda)$ and variance $\sigma^{2}$~\cite{queuingtheory}. There are two types of devices~(servers) with mean service rates $\mu_{1}$ and $\mu_{2}$ respectively. We have ignored cold misses and analyzed using capacity misses with evictions.

\begin{equation}
T_{h_i} = n_{i_r}*\frac{1}{\mu_{1_r}}+n_{i_w}*\frac{1}{\mu_{1_w}}, \forall 1\leq i\leq P
\label{eq:eq1}
\end{equation}

\begin{equation}
T_{m_i} =  n_{i_{m}}*\frac{1}{\mu_{2}}
\label{eq:eq2}
\end{equation}

\begin{equation}
T_i = \max(T_{h_i},T_{m_i}), \forall 1\leq i\leq P      
\label{eq:eq3}
\end{equation}

\begin{equation}
T = \max(T_i), \forall 1\leq i\leq P
\label{eq:eq4}
\end{equation}

The total service time for the two-tier storage system for a workload with $n_{i_r}$ read requests and $n_{i_w}$ write requests per process is described using equations~\ref{eq:eq1} to \ref{eq:eq4}. We have used service rates $\mu_{1_r}$ and $\mu_{1_w}$ for tier 1 read and write hits respectively. These values can be computed using the performance models for NVMes~(subsection~\ref{subnv}) and RPC communication costs. $T_{h_i}$ is the minimum time for servicing hits by process $i$ using $k$ RPC threads per process. Let $n_{i_m}$ be the number of capacity misses per process. Let $\mu_{2}$ be the miss service rate. The values of $\mu_{2}$ can be computed using the performance models for parallel IO using HDDs~(subsection~\ref{subhd}). $T_{m_i}$ is the miss penalty per process. Since the IO thread and RPC service threads execute concurrently, $T_i$, the total time for process $i$ to empty all queues is defined by equation~\ref{eq:eq3}. The total time across all processes is the maximum $T_i, \forall 1\leq i\leq P$, where $P$ is the number of processes.

But, equations~\ref{eq:eq1} to \ref{eq:eq4} do not model other quantities of interest such as response rates and waiting times in the tiers. Let $\lambda$ be the rate at which read/write requests are generated by a workload. To simplify equations, let $E(\mu_1)$ and $E(\mu_2)$ be the expected service rates of tiers 1, 2 and let $p_{12}$ be the miss rate of the workload. The model description provided here is for a single process.
The queuing network model of the system depends on its implementation. If hits and misses are serviced by the same set of $k$ threads per process, then the system can be modeled using a single $M/G/k$ queue, where arrivals from two populations enter the system with arrival rates $\lambda*p_{12}$ and $\lambda*(1-p_{12})$. These two types of requests have different service times $\frac{1}{E(\mu_1)}$ and $\frac{1}{E(\mu_2)}$. The expected service time for the system is provided in equation~\ref{eq:eqn5}.
\begin{equation}
\frac{1}{\mu} = (1-p_{12})*\frac{1}{E(\mu_1)}+p_{12}*\frac{1}{E(\mu_2)}
\label{eq:eqn5}
\end{equation}	

The expected arrival rate is denoted by $\lambda$. The utilization ratio of hits is $\rho_1 = \frac{(1-p_{12})*\lambda}{E(\mu_1)}$ and misses is $\rho_2 = \frac{p_{12}*\lambda}{E(\mu_2)}$. The utilization ratio for the system is $\rho = \frac{\lambda}{\mu}$. The number of requests in service and waiting in the queue can be computed using the values of arrival rates, service rates and utilization ratios for each population and also for the entire system~\cite{queuingsystem}.  
The implementation described in this paper uses a separate IO thread for page misses, refer to figure~\ref{fig:fig5}. In this case, an $M/G/k$ queue is used for servicing hits and an $M/M/1$ queue for misses. The arrival and service rates of the miss queue are $p_{12}*\lambda$ and $E(\mu_2)$ respectively. On exiting the miss queue, requests enter the $k$-server queue. Therefore, the $k$-server queue has two types of traffic entering it at different rates ; hits at rate $(1-p_{12})*\lambda$ and misses at rate $E(\mu_2)$ from the single server queue. Requests exit the system via the $k$-server queue.
The effective arrival rate at the $k$-server queue is $(1-p_{12})*\lambda+E(\mu_2)$. The utilization ratio of the $k$-server queue is $\rho_1 = \frac{(1-p_{12})*\lambda+E(\mu_2)}{E(\mu_1)}$. The utilization ratio of the IO queue is $\rho_2 = \frac{p_{12}*\lambda}{E(\mu_2)}$.  

This system can be analyzed at equilibrium to compute values of expected queue lengths~($L_{1}$,$L_{2}$) and waiting times~($W_{1}$,$W_{2}$) per process for the queues using equations~\ref{eq:eq6} and \ref{eq:eq7}. Equation~\ref{eq:eq6} describes an $M/M/k$ queue with the same expected service rate for reads and writes, where $P_{0}$~(probability of the queue being empty) can be computed using the full $M/M/k$ model described in \cite{queuingsystem}. Similar equations can be derived for an $M/G/k$ queue using the mean and variance of the read/write service~(hit) time distribution. Equation~\ref{eq:eq7} describes an $M/M/1$ queue for IO misses.  

\begin{equation}
L_{1} = \frac{P_{0}*\rho_{1}^{(k+1)}}{(k-1)!(k-\rho_{1})^2},
W_{1} = \frac{L_{1}}{(1-p_{12})*E(\lambda)+E(\mu_2)}
\label{eq:eq6}
\end{equation}
\begin{equation}
L_{2} = \frac{\rho_{2}^{2}}{(1-\rho_{2})},
W_{2} = \frac{L_2}{E(\lambda)*p_{12}}
\label{eq:eq7}
\end{equation}

The analysis using queuing networks is useful when the system is at equilibrium, i.e. all utilization ratios are $<1$. The configuration of the two-tier system can be adjusted for desired service rates while maintaining the equilibrium of the system. If not in equilibrium, equations~\ref{eq:eq1} to \ref{eq:eq4} can be used to estimate minimum execution times.
For example, consider a workload of $n=10000$ read requests distributed over $2000$ pages~($page size=524288$ bytes) run using $4$ MPI processes. Suppose each process has a $32GB$ tier 1 cache. Assume the pages and requests are uniformly distributed across processes. Let $p_{12}=0.2$ be the miss rate per process. Let $\lambda=100$reqs/sec, $\mu_{1}=1000$reqs/sec and $\mu_{2}=33$reqs/sec. The value of $\mu_{1}$ is lower than NVMe throughput rates because it includes RPC and synchronization costs. The effective arrival rate $\lambda = 0.8*100+0.2*0.33 = 86.6$reqs/sec. The utilization ratios $\rho_{1} = \frac{86.6}{1000} = 0.0866$ and $\rho_{2} = \frac{20}{33} = 0.6$. This system is in equilibrium, because the utilization ratios of both queues are $<1$. The expected length of the tier 1 queue is almost $0$. The number of requests per process = $2500$ and the number of misses per process = $500$. The expected time taken for $2500$ arrivals is $\lambda*T=2500$, $86.6*T=2500$, $T=28.8$ sec. The total response time per process = $\frac{2500}{1000}=2.5$ sec. We haven't included RPC communication costs in our performance models, but they can be computed from LogP communication model~\cite{logp} or by profiling RPC benchmarks.  
For the analysis, we have assumed that the request address stream is random. If a stream contains consecutive addresses to the same page, they can be grouped into a single request and forwarded to the miss queue. This will increase the mean service rate of the miss queue. 
We have assumed that all processes have the same request arrival and miss rates and that all NVMes are the same. If arrival rates and the probability distributions of requests to processes~(mapping) are known, then the input distributions of processes can be modeled using separate random variables. If the load is not equally distributed, then processes will have different miss rates. Since HDDs are shared, $\mu_{2}$ is a global variable. The $M/G/k$ queues have to be replaced by $G/G/k$ queues for such general cases. Quantities such as expected queue lengths and waiting times will be different for each process and the mean or maximum values may be chosen as system-wide global values.   
Harder quantities to model are $n_{m_i}$, the number of misses because it depends on workloads and cache sizes.
All the performance models in this section were built using linear regression~\cite{statistics}. We used R~\cite{Rmanual} for modeling and analysis. We used cross-validation~\cite{statistics} to reduce over-fitting. All training experiments were performed on the Delta Supercomputer\cite{delta}. Delta is described in detail in the evaluation section~\ref{6}.

\subsection{Solid-State Devices~(NVMes)~\label{subnv}}
In this section, we describe models for predicting the performance of NVMes from a set of parameters. We used NVMes by mapping posix files to virtual memory. Read/write operations on these files are transferred over the network~(PCIe here) and submitted to concurrent IO queues on the NVMes. Let $n$ be the number of read/write requests issued to a file stored on an NVMe. NVMes can be modeled using multi-server $G/G/k$ queues~\cite{queuingtheory}. Using NVMes via page mapping adds overheads from page lookups and page faults. Highly correlated read requests benefit from mapping because they have fewer NVMe accesses. To utilize the concurrency provided by NVMes, they should be used directly by submitting requests asynchronously to their IO queues. Libraries such as libaio~\cite{libaio} or spdk~\cite{spdk} can be used for direct access of NVMes. These libraries also provide APIs for high throughput communication protocols such as RDMA for transferring requests~\cite{nvperfr}. Our performance models do not include the communication protocol or the mode of use of NVMes as parameters. Therefore, the same model can be trained for different systems.
The total cost of a set of IO operations on a file stored on an NVMe depends on the mean request size, communication costs, NVMe configuration, workload size and the concurrency in the workload. The concurrency in a workload depends on the request type~(read/write), number of IO queues in use, number of requests, total address range in use and the request generation rate. 

We identified the following parameters to model the total time (training set is provided in brackets):
\begin{itemize}
\item Number of client threads~($X_{1}$) : $[8,16,32,64]$
\item Number of distinct block addresses~($X_{2}$) 
\item IO request size~($X_{3}$) : $[512,4096,8192,65536,262144]$
\item Number of IO requests~($X_{4}$) : $[1000-4000000]$
\item Total address range in use~($X_{5}$) : $500MB-500GB$
\end{itemize}

Let $Y$ be the total read/write time for any workload. Our performance model is provided by equation~\ref{nvmeperfen}.
\begin{equation}
Y = X_1*X_3*X_4+X_5*X_4*X_3
\label{nvmeperfen}
\end{equation}	

\begin{table}[!tbph]
\scriptsize
\begin{tabular}{|c|c|c|c|c|}
\hline
  & Estimate & Std. Error& t value & $Pr(>|t|)$ \\
\hline
(Intercept)& -5.941e+00 & 1.560e+01 & -0.381 &  0.70353\\
\hline
x1   &6.252e-01 & 4.387e-01 &  1.425 & 0.15490\\
\hline
x3   &-6.326e-05 & 2.143e-04 & -0.295 & 0.76801\\
\hline
x4   &3.726e-05 & 1.860e-05 &  2.003 & 0.04580 \\
\hline
x5   &6.213e-11&  5.174e-11 &  1.201 & 0.23053\\
\hline
x1:x3  &1.667e-06 & 6.784e-06 &  0.246 & 0.80598\\
\hline
x1:x4  &-8.464e-07 & 5.005e-07 & -1.691 & 0.09158 \\
\hline
x3:x4  &-1.650e-09 & 5.655e-10 & -2.917 & 0.00373 \\
\hline
x4:x5  &2.029e-16 & 8.570e-17 &  2.368 & 0.01834 \\
\hline
x3:x5  &-6.564e-16 & 1.541e-15 & -0.426 & 0.67030\\
\hline
x1:x3:x4 &1.973e-10 & 1.510e-11 & 13.061 & $< 2e-16$\\
\hline
x3:x4:x5 &1.103e-20 & 2.343e-21 &  4.706 & 3.46e-06\\
\hline
\end{tabular}
\vspace{1mm}
\caption{NVMe Write Performance Model~\label{tab:tab1w}}
\end{table}

We trained separate performance models for reads and writes because read/write bandwidths are different. Our performance model includes singleton terms as well as interactions between parameters. Our hypothesis is that interactions between parameters are useful for modeling load distributions and concurrency in the performance models of hardware devices and concurrent software~\cite{perfmodelgpu}. The write performance model is tabulated in table~\ref{tab:tab1w} with parameters and significance~(column $Pr(>|t|)$) values. Lower the values of $(Pr(>|t|)$, higher the significance of the corresponding terms in the performance model. The significant terms of this model are $X_1:X_3:X_4$ and $X_3:X_4:X_5$ which have the least probabilities. The total write time depends on the number of client threads, IO request size and the number of requests because these parameters~($X_1:X_3:X_4$) model the load distribution. The term $X_3:X_4:X_5$ models page faults and NVMe write costs such as garbage collection~\cite{gc}. Since the NVMe configuration is fixed, garbage collection intervals depend on the number of blocks in use, which can be modeled using $X_5$, $X_4$ and $X_3$. Concurrency of the workload is captured by $X_4:X_5$, i.e. the total number of requests and their address range. Terms which did not have affect on the model are $X_1$, $X_3$ and $X_5$ in isolation along with $X_1:X_3$ and $X_3:X_5$. Request size and number of client threads affect write performance only when combined with $X_4$, i.e. the total number of requests which is evident from the observations of probability values in table~\ref{tab:tab1w}.

\begin{figure}[!tbph]
\begin{center}	
\includegraphics[width=2.2in,height=2.2in]{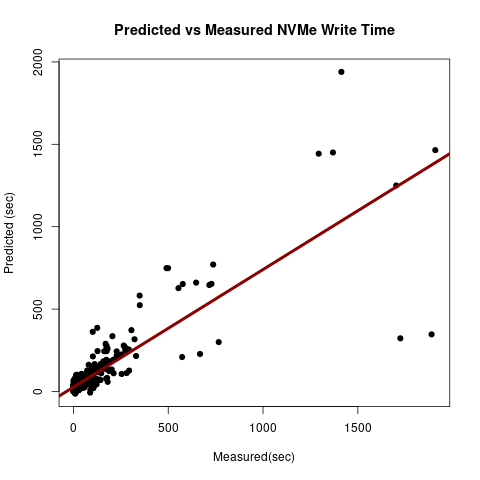}	
\caption{NVMe Write Performance Model Fit~\label{fig:fig6}}
\end{center}
\end{figure}

This model captures the effects of dependent parameters on the output accurately and verifies our initial hypothesis about NVMe write performance. It was trained using ~400 experiments. The cross-validation parameter was $K=20$. The AIC score for the linear regression model was 5267.4 and the goodness of fit is shown in figure~\ref{fig:fig6}. We chose the best model after comparing it with similar models using Anova~\cite{Rmanual}.

\begin{table}[!tbph]
\scriptsize
\begin{tabular}{|c|c|c|c|c|}	
\hline
& Estimate &Std. Error &t value &$Pr(>|t|)$ \\
\hline
(Intercept)& -6.059e+00 & 8.802e+00 & -0.688 & 0.491565\\
\hline
x1 & 2.182e-02 & 2.475e-01 & 0.088 & 0.929812\\
\hline
x3 & 1.009e-04 & 1.209e-04 &  0.835 &0.404440\\
\hline
x4 & -3.566e-06 & 1.049e-05 & -0.340 &0.734131\\
\hline
x5 & 6.963e-11 & 2.920e-11 &  2.385 &0.017533 \\
\hline
x1:x3 &-2.066e-07 & 3.828e-06 & -0.054 &0.956978\\
\hline
x1:x4 &-1.165e-08 & 2.824e-07 & -0.041 &0.967125\\
\hline
x3:x4 &-4.060e-10 & 3.191e-10 & -1.272 &0.203981\\
\hline
x4:x5 &1.259e-16 & 4.835e-17  & 2.603 &0.009572\\
\hline
x3:x5 &-2.984e-15 & 8.693e-16 & -3.433 &0.000658\\
\hline
x1:x3:x4 &-6.675e-12 & 8.522e-12 & -0.783 &0.433929\\
\hline
x3:x4:x5 &1.896e-20 & 1.322e-21 & 14.340 & $< 2e-16$\\
\hline
\end{tabular}
\vspace{1mm}
\caption{NVMe Read Performance Model\label{tab:tab1r}}
\end{table}

\begin{figure}[!tbph]
\begin{center}
\includegraphics[width=2.2in,height=2.2in]{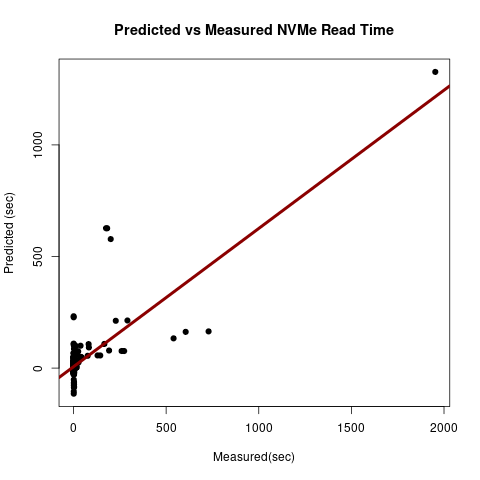}
\caption{NVMe Read Performance Model Fit~\label{fig:fig7}}
\end{center}
\end{figure}

We used the same set of parameters for both read and write performance models. The performance model for reads was also selected after comparing with other models using Anova. The terms of the read performance model are tabulated in table~\ref{tab:tab1r}. Unlike writes, read operations have no contention. Similar to writes, the work per thread is modeled using $X_3:X_4:X_1$ and page faults are modeled by terms containing $X_{5}$. $X_1$ has lower significance in this model compared to the write performance model, because of page mapping and zero contention. Pages that can be stored in page tables are reused during reads. Terms containing $X_3$ have higher significance because they affect page boundaries and page faults, which lead to NVMe accesses. The AIC score of this model was 4786.4. The goodness of fit of the read performance model is provided in figure~\ref{fig:fig7}. We found regression to be a useful technique for modeling load distribution, contention and hardware features such as block garbage collection in NVMes, memory controllers and page table sizes. It was also a useful tool for inferring the significance of terms and for verifying hypotheses about expected behavior.    
We did not parameterize request rate in these models. The ratio of reads to writes may be added as a parameter to create a single model. Instead we chose to create separate models and determine the performance of a mixed workload by adding individual costs. The actual cost of a mixed workload is likely to be lower than the sum. If NVMes are accessed directly instead of page mapping, the relative significance of terms in these models will change. $X_1$ may be replaced by the number of IO queues in this case. Our observations about the write performance model are not likely to differ. The read performance model is likely to have different significance values for its terms because of the absence of page tables.
\subsection{Hard Disk Drives~(HDDs)\label{subhd}} 
Performance models for shared HDDs were built by the HPC community for exploring file layouts that minimized IO time. The objectives in those models were to determine the best file layout~(stripe size, stripe count) that minimized IO time for an application, given a certain process count~\cite{autotuning}. There have been other efforts to model IO performance overheads and variance using machine learning~\cite{autotuning},~\cite{mlhdd1},~\cite{mlhdd2},~\cite{mlhdd3} using data center workloads. Our objective was to model the total time for reading/writing a file, given its layout, and to compute the mean read/write time per stripe from total time. The costs of cache misses, evictions and prefetches can be computed as functions of the number of stripes. Separate models were created for reads and writes because of bandwidth differences. We chose the following parameters and training set values to model the total IO access time :

\begin{itemize}
\item Number of processes~($X_1$) : $[4,8,16,32,64,128,200]$
\item Number of disks~(Stripe count $X_2$) : $[1,2,4,8]$
\item Number of stripes per disk~($X_3$) 
\item Stripe size~($X_4$) : $[64KB-64MB]$
\item File size~($X_5$)	: $[100MB-350GB]$
\end{itemize}		

Let $Y$ be the total read/write time for a file. Our performance model for hard disk drives is provided by equation~\ref{diskperfmodel}.
\begin{equation}
Y = X_3*X_4+X_5*X_1*X_2
\label{diskperfmodel}
\end{equation}

\begin{table}[!tbph]
\scriptsize
\begin{tabular}{|c|c|c|c|c|}
\hline
           &   Estimate & Std. Error & t value &$Pr(>|t|)$\\
\hline
(Intercept) & 7.297e+00 & 5.837e+01 &  0.125 & 0.90066\\
\hline
x3  &4.318e-04 & 1.776e-04 &  2.432 & 0.01605 \\
\hline
x4  &-4.354e-06 & 1.464e-06 & -2.974 & 0.00336\\
\hline
x5  &1.002e-08 & 1.321e-09 &  7.586& 1.90e-12 \\
\hline
x1 &3.869e-01 & 8.273e-01 &  0.468 & 0.64059\\
\hline
x2  &6.664e+00 & 1.060e+01 &  0.629 & 0.53045\\
\hline
x3:x4 &2.007e-11 & 1.820e-09 &  0.011 & 0.99122\\
\hline
x5:x1 &-7.486e-11 & 1.208e-11 & -6.196 &4.07e-09\\
\hline
x5:x2 &-9.269e-10 & 2.033e-10 & -4.560 &9.61e-06 \\
\hline
x1:x2 &-9.916e-02 & 1.444e-01 & -0.687 & 0.49310\\
\hline
x5:x1:x2  & 8.344e-12 & 1.890e-12 &  4.416 &1.76e-05\\
\hline
\end{tabular}
\vspace{1mm}
\caption{HDD Write Performance Parameters~\label{tab:tab3w}}
\end{table}

\begin{figure}[!tbph]
\begin{center}
\includegraphics[width=2.2in,height=2.2in]{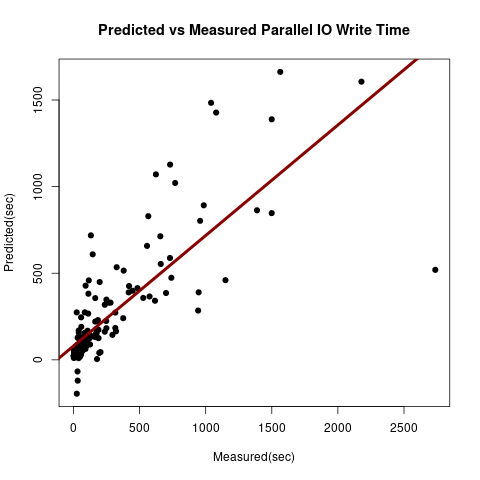}
\caption{Parallel IO Write Performance Model for HDD fitness\label{fig:fig8}}
\end{center}
\end{figure}

We used observations from ~200 separate experiments for reads and writes. Each experiment accessed the entire file once in parallel. 
The best models were chosen after comparing with similar models using Anova. We used interaction terms to model IO request distribution on disks and data transfer costs between processes and disks. We used parallel HDF5 on Lustre file system for these experiments. The terms of the write performance model are tabulated in table~\ref{tab:tab3w}. Stripe count~($X_2$), stripe size~($X_4$) and number of requests per disk~($X_3$) were found to be significant factors in determining the parallel IO time. HDF5 IO requests are broken down into Lustre requests not exceeding stripe size. These requests are queued and serviced independently by disks. Since we accessed the entire file, $X_5:X_2$ and $X_3:X_4$ model the load per disk. Between the two terms, $X_5:X_2$ models the entire load as a single large contiguous request, while $X_3:X_4$ treats the load as a function of the number of requests and stripe size. In our model described in table~\ref{tab:tab3w}, $X_5:X_2$ was significant, but $X_3:X_4$ had low significance. We used $X_5:X_1:X_2$ to model the total communication time for transferring $X_5$ bytes by $X_1$ processes to $X_2$ disks or vice versa. $X_5:X_1$ models the communication cost incurred by $X_1$ processes. $X_1:X_2:X_5$ has more significance compared to $X_1:X_2$ because it incudes the total size of data transferred over the network. Depending on the file size, data may be transferred between processes and disks as variable length messages over multiple iterations. Therefore, we modeled the communication cost of the entire data transfer instead of adding a parameter for message size. 
The goodness of fit is provided in the figure~\ref{fig:fig8} and the AIC score for this model was 2566.5. 

\begin{table}[!tbph]
\scriptsize
\begin{tabular}{|c|c|c|c|c|}
\hline
&        Estimate& Std. Error& t value& $Pr(>|t|)$\\
\hline
(Intercept)& -3.771e-01 & 8.013e+01 & -0.005&  0.99625\\
\hline
x3 &5.913e-04 & 2.106e-04 &  2.808 & 0.00573 \\
\hline
x4 &-1.584e-06 & 1.729e-06 & -0.916 & 0.36136\\
\hline
x2  &8.933e+00 & 1.326e+01 &  0.673 & 0.50180\\
\hline
x1  &-2.563e+00 & 1.400e+00 & -1.830 & 0.06944 \\
\hline
x5  &6.274e-10 & 2.154e-09 &  0.291 & 0.77131\\
\hline
x3:x4 &1.715e-08 & 2.718e-09 &  6.312 &3.75e-09 \\
\hline
x2:x1 &3.694e-01 & 2.113e-01 &  1.749 & 0.08262 \\
\hline
x2:x5 &-2.272e-10 & 2.550e-10 & -0.891 & 0.37452\\
\hline
x1:x5 &-4.751e-11 & 2.038e-11 & -2.332 & 0.02121 \\
\hline
x2:x1:x5 &5.167e-12 & 2.662e-12 &  1.941 & 0.05435 \\
\hline
\end{tabular}
\vspace{1mm}
\caption{HDD Read Performance Model Parameters~\label{tab:tab3r}}
\end{table}

\begin{figure}[!tbph]
\includegraphics[width=2.2in,height=2.2in]{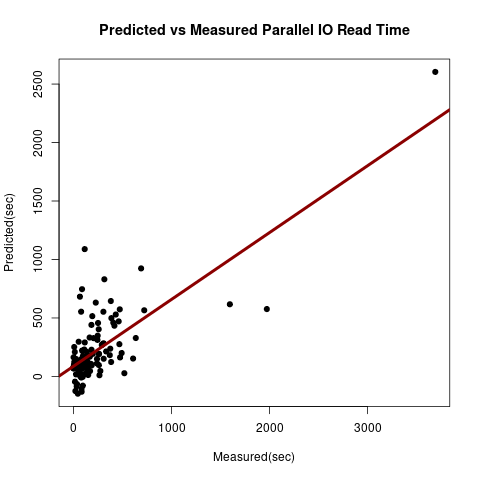}
\caption{Parallel IO Read Performance Model for HDD fitness~\label{fig:fig9}}
\end{figure}

The same parameters were used to model the parallel read performance of HDDs. The terms of the read performance model and their significance are tabulated in table~\ref{tab:tab3r}. The read model terms have different significance values. Both $X_5:X_2$ and $X_3:X_4$ are significant terms in this model. The differences between the two models for these terms is likely to be caused by insufficent training sets in terms of size and range. The AIC score of this model was 2035.1 and its goodness of fit is provided in figure~\ref{fig:fig9}. 

\section{Evaluation\label{6}}
All experiments described in this paper were performed on the Delta Supercomputer~\cite{delta} at the National Center for Supercomputing Applications~(NCSA). Delta has AMD Milan CPUs~(8 NUMA nodes, 64 cores). Each CPU node has a local NVMe~(0.7TB) attached to it via PCIe. Delta has Lustre PFS with upto 8 disks capable of 6PB storage. Delta has 128 CPU nodes in total. The communication network is high-speed cray slingshot~\cite{slingshot}. The two-tier storage software was implemented in C++ using GCC and MPI. The other libraries we used are Mercury~\cite{mercury} for multi-threaded RPCs, Intel-TBB~\cite{tbb} for memory allocation, read-write locks and concurrent data structures, Intel-PMDK~\cite{pmdk} for mapping NVMe posix files to CPU memories and HDF5 for parallel IO~\cite{hdf5}. The NVMes were used exclusively, while the HDDs were shared with other users on the cluster. 
\subsection{Online Learning}
We used two IO traffic models to test the weight-sharing cache replacement algorithm. The traffic models used are \emph{Poisson}~\cite{queuingtheory} and \emph{IRM}~\cite{lfumodel}. In the Poisson model, the probability of a page request decreases exponentially with time since its arrival. It defines the temporal locality of a page in the workload. We chose suitable Poisson functions to ensure that the temporal locality of the workload is slow evolving~\cite{belady}. In our traffic models, spatial locality is defined w.r.t. to a page. In the Independent Reference Model~(IRM), pages have fixed lifetimes and popularities~(maximum requests). The popularity distribution of pages follows a Zipf distribution~\cite{zipf}. A page expires when its number of requests have exceeded its allowed maximum. Most IO traffic fall into one of these two models. Once a page is fetched into the cache, its reuse depends on its lifetime or popularity. In the Poisson model, pages with longer lifetimes have higher chances of reuse while in the IRM model, majority of IO requests are distributed across the most popular pages. IRM workloads can have sharp changes in temporal locality. IO accesses common in HPC workloads~\cite{iopatterns} is closest to a Poisson model with the same lifetimes for all pages. We chose these two traffic models because their miss streams differ for LRU and LFU. In a Poisson model, LRU evicts pages with expired lifetimes with high probability. LFU is more suitable for the IRM model, because it evicts least popular pages with high probability. We compared the number of cache misses using both types of traffic models for LRU, LFU and weight-sharing online learning~(WS) cache replacement policies. The observations from the Poisson model and the IRM model experiments are tabulated in tables~\ref{tab:table1} and ~\ref{tab:table2}. We used 1 MPI process, $64$ cache lines with line size $8192$ bytes for these experiments. The results for LRU and LFU in tables~\ref{tab:table1} and ~\ref{tab:table2} match our speculations about the temporal locality of these models. The WS replacement algorithm could learn the traffic models and switch between experts. The number of cache misses using WS is comparable to LRU for Poisson traffic and LFU for IRM traffic. In some cases, WS performed better than both because it could adapt to variations by choosing the most appropriate expert at a particular instant in time. The time taken by WS for OL decisions is presented in the tables under the column labeled \textbf{WS~(sec)}.
We have used OL to learn the IO traffic from past accesses and restricted ourselves to LRU and LFU. It has been shown that these algorithms perform as well as clairvoyant \emph{MIN}~(within a constant factor)~\cite{tarjan}. The OL algorithm can be extended by adding a \emph{farthest-in-future} expert for improving cases which have easily identifiable sequences. We found WS to be a low overhead cache replacement technique suitable for IO workloads.
\begin{table}[!tbph]
\tiny	
\begin{center}
\begin{tabular}{cc}
\begin{minipage}{0.5\linewidth}
\begin{tabular}{|c|c|c|c|}
\hline
\textbf{\#reqs}&\textbf{LRU}&\textbf{LFU}&\textbf{WS~(sec)}\\
\hline
500&252&220&206~(0.000684891)\\
\hline
1000&396&410&390~(0.00156582)\\
\hline
2500&1018&1030&951~(0.00437329)\\
\hline
5000&1991&2034&1907~(0.299254)\\
\hline
10000&3871&4128&3927~(0.627381)\\
\hline
\end{tabular}
\vspace{1mm}
\caption{Poisson IO traffic Model\label{tab:table1}}
\end{minipage}
\begin{minipage}{0.5\linewidth}
\begin{tabular}{|c|c|c|c|}
\hline
\textbf{\#reqs}&\textbf{LRU}&\textbf{LFU}&\textbf{WS~(sec)}\\
\hline
500&377&393&376~(0.00149009)\\
\hline
1000&763&731&735~(0.00317623)\\
\hline
2500&1918&1833&1841~(0.00835657)\\
\hline
5000&3762&3676&3673~(0.0170301)\\
\hline
10000&7504&7121&7138~(0.03388)\\
\hline
20000&14684&13851&13899~(0.0661851)\\
\hline
\end{tabular}
\vspace{1mm}
\caption{IRM IO traffic Model\label{tab:table2}}
\end{minipage}
\end{tabular}
\end{center}
\end{table}

\subsection{Performance}
In this section we evaluate the parallel performance of the storage benchmark using different workloads. We tested the full implementation~(including OL eviction algorithm and stride identifiers) for communication bottlenecks in request processing at scale.

\begin{figure}[!tbph]
\begin{center}	
\includegraphics[width=3in,height=2.2in]{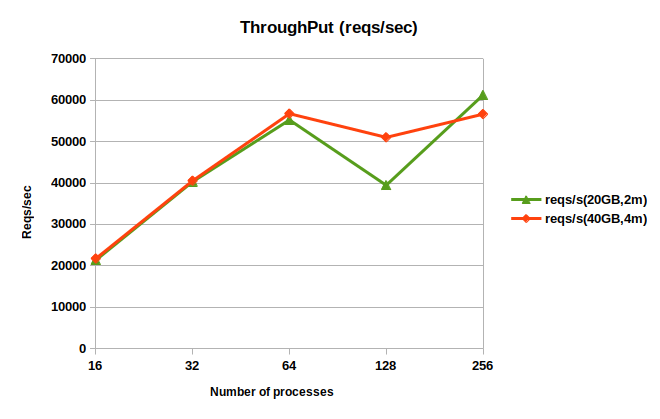}
\caption{Read Throughput Vs Number of Processes\label{fig:fig10}}
\end{center}
\end{figure}

The graph in figure~\ref{fig:fig10} shows the throughput of the two-tier storage system for a read workload with increasing number of MPI processes. The input file size was 500GB. This workload has two test cases :
\begin{itemize}
\item $2$million $128$-byte read requests distributed over 20GB
\item $4$million $128$-byte read requests distributed over 40GB
\end{itemize}
Each MPI process had $32GB$ cache allocated on tier 1 NVMes. We placed $4$ MPI processes per node and used $64$ threads per process to service RPC requests. $16$ client threads were spawned per process for submitting requests. The request arrival rate was $100$reqs/sec. The page size and stripe size were $524288$ bytes for these experiments and stripe count was $8$. These experiments used block partitioning to partition pages across caches. Read requests were processed by accessing data from local NVMes without inter process communication. From the observations in the graph in figure~\ref{fig:fig10}, throughput increases with increasing processes. The drop in performance is due to page misses which may be affected by congestion in the network or by interference from other IO workloads on the cluster.

We performed strong scaling experiments using random mapping and read/write workloads to understand the inter process communication costs of multi-threaded remote RPCs and also to evaluate dependencies between the quantities defined in equation~\ref{eq:eq3} and response time.
The two workloads used are described below :
\begin{enumerate}
\item Workload1 : This is a low reuse workload, with $5$million requests distributed over $229376$ pages. Page size was $524288$ bytes and request size was $512$ bytes. The workload accessed approximately $100GB$ data from a $400GB$ file. The request arrival rate was $100$reqs/sec. This workload and cache configuration includes systems that are not in equilibrium.  
\item Workload2 : This is a high reuse workload, with $8$million requests spread over $32768$ pages. Page size was $524288$ bytes. It used approximately $20GB$ data from a $400GB$ file. The request arrival rate was $100$reqs/sec. 
\end{enumerate}

Each MPI process has $32GB$ cache allocated on tier 1 NVMes. We placed $4$ MPI processes per node and used $16$ threads per process to service RPC requests. $16$ client threads were spawned per process for submitting requests. Stripe count was $8$ and the stripe size was equal to page size for both workloads.

\begin{table}[!tbph]
\scriptsize
\begin{tabular}{|c|c|c|c|}
\hline
\#procs& response time(s)& Mean Idle Time(s) &HDD time(s)\\
\hline
16&740.063 &1.12738& 637.148 \\
\hline
32 &415.31 &1.47583& 373.875 \\
\hline
64 & 398.438&3.0563&447.962\\
\hline
128  &353.945 &4.90604&473.151\\
\hline
200 &444.919 &12.6391& 989.649\\
\hline
\end{tabular}
\vspace{1mm}
\caption{Strong Scaling Performance of Two-tier Storage for Read-Modify-Write Workload1~\label{tab:tab5}}
\end{table}

\begin{figure}[!tbph]
\includegraphics[width=2.8in,height=2.2in]{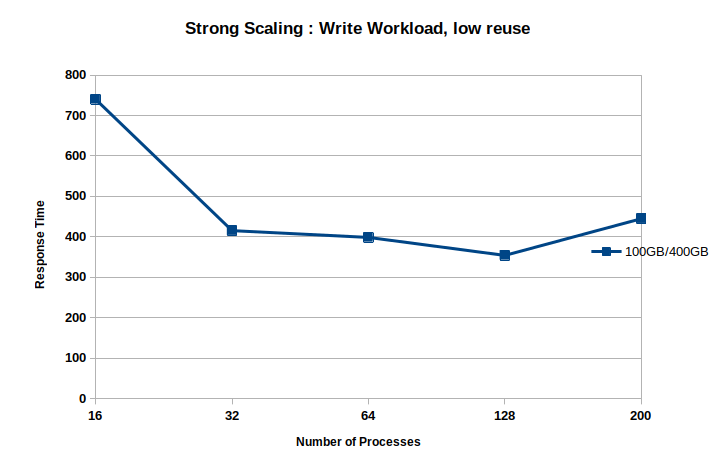}
\caption{Strong Scaling of Two-tier Storage for Read-Modify-Write Workload1~\label{fig:fig11}}
\end{figure}

Table~\ref{tab:tab5} has tabulated the observations for Workload1 with read-modify-write requests. Response time was measured as the maximum time taken to complete request processing across all processes and client threads. The response time scales with increasing number of processes until $128$ processes. Performance dropped at $200$ processes because of the high IO miss time~(HDD). The communication time of IO misses depends on the cluster design, such as number of IO servers, HDDs and network topology. This workload could cause congestion because there may be several large messages~(stripe size) in transit on the network. Miss service time also depends on the interference caused by other workloads sharing HDDs. In this experiment, $T_{m_i} > T_{h_i}$~(equation~\ref{eq:eq3}) and the completion time~($\max(T_{h_i},T_{m_i}$) is mostly dominated by miss penalty. The strong scaling graph of these experiments is plotted in the figure~\ref{fig:fig11}.

\begin{table}[!tbph]
\scriptsize
\begin{tabular}{|c|c|c|c|}
\hline
\#procs& response time(s)& HDD time(s)& completion time(s)\\
\hline
16& 509.089 & 70.9043 & 509.089 \\
\hline
32 &254.459 &51.3207  & 254.459\\
\hline
64 & 193.333& 96.5852 & 193.333\\
\hline
128  &140.96 & 102.139 & 140.96 \\
\hline
200  & 103.914& 78.7909 & 103.914\\
\hline
512 & 44.051 & 36.574& 44.051\\ 
\hline 
\end{tabular}
\vspace{1mm}
\caption{Strong Scaling Performance of Two-tier Storage with Read-only Workload2~(4KB requests)~\label{tab:tab6}}
\end{table}

\begin{table}[!tbph]
\scriptsize
\begin{tabular}{|c|c|c|c|}
\hline
\#procs& response time(s)& HDD time(s)& completion time(s)\\
\hline
16& 841.333 & 53.9377 & 841.333 \\
\hline
32 &438.327 &54.6357  & 438.327\\
\hline
64 & 253.772& 50.4423 & 253.772\\
\hline
128  &299.586 & 319.854 & 319.854 \\
\hline
256  & 201.257& 195.658 & 201.257\\
\hline
512 & 87.4437& 174.346& 174.346\\
\hline
\end{tabular}
\vspace{1mm}
\caption{Strong Scaling Performance of Two-tier Storage with Read-Modify-Write Workload2~(64KB requests)~\label{tab:tab7}}
\end{table}

\begin{figure}[!tbph]
\includegraphics[width=3in,height=2.2in]{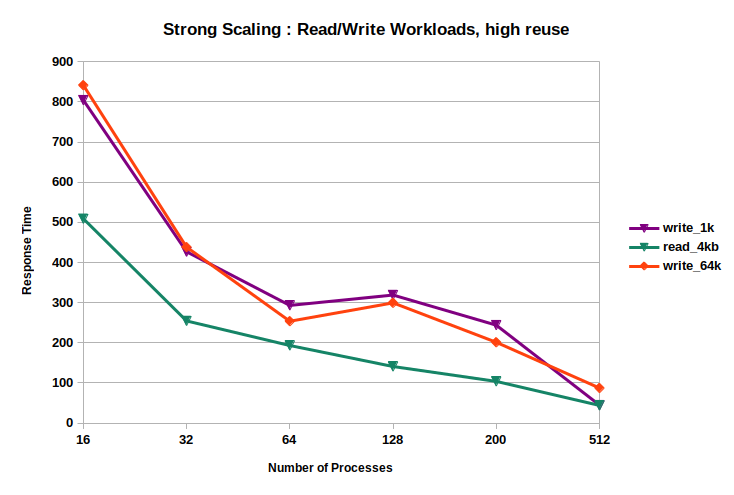}
\caption{Strong Scaling Performance of Two-tier Storage with Read/Read-Modify-Write Workloads2~\label{fig:fig12}}
\end{figure}

From the observations in tables~\ref{tab:tab6} and \ref{tab:tab7} and figure~\ref{fig:fig12}, we find that workload2 scales strongly with increasing number of processes, because $T_{h_i} > T_{m_i}$. We ran three experiments using workload2, one read workload with 4KB requests, and two read-modify-write workloads with 1KB and 64KB requests. Writes are more expensive than reads for the same number of hits and misses. Although the test cases accessed data from random processes in the cluster we do not observe sharp increase in response time due to inter process communication or runtime overheads in the software implementation. The high-speed slingshot network on Delta is one of the reasons for strong scaling of short messages in spite of the high volume of inter process communication. These experiments were used to evaluate the software design. To summarize, tiering benefits both types of workloads, but scalability is a challenge for workload1. For workload1, one option to improve scalability is to reduce the request arrival rate which will decrease the values of $\rho$ and also request queue lengths at both tiers. If that is not possible, another option is to increase the unit of data transfer between tiers 1 and 2. A real application will have computation kernels in addition to IO which will reduce the mean arrival rates of IO requests.  
\section{Conclusions and Future Work\label{7}}
We could identify parameters that affect the performance of a tiered storage system. For required arrival and service rates, these performance models can be used to configure cache size~(miss rate), number of processes and data sizes at each tier. We found it important to model the behavior of the system for supporting applications with different requirements, e.g. high throughput GPU applications and slower checkpointing. To conclude, tiering is a useful design choice for storage systems in large clusters, refer to IO communication times for large messages, network congestion and server load imbalance. The storage devices may be homogeneous or mixed. We would like to improve the performance models, the behavioral models of devices as well as evaluate prefetching. Other directions for future work are to design learning algorithms for data migration between caches w.r.t. IO request distribution. It is also worthwhile to investigate IO prefetchers which do not require offline training.

\section{Acknowledgements\label{8}}
This research was funded by DOE grant \#DE-SC0023263 and used the Delta advanced computing and data resource which is supported by the National Science Foundation (award \#OAC 2005572) and the State of Illinois.

\bibliographystyle{IEEEtran}
\bibliography{references}

\begin{thebibliography}{10}
\providecommand{\url}[1]{#1}
\csname url@samestyle\endcsname
\providecommand{\newblock}{\relax}
\providecommand{\bibinfo}[2]{#2}
\providecommand{\BIBentrySTDinterwordspacing}{\spaceskip=0pt\relax}
\providecommand{\BIBentryALTinterwordstretchfactor}{4}
\providecommand{\BIBentryALTinterwordspacing}{\spaceskip=\fontdimen2\font plus
\BIBentryALTinterwordstretchfactor\fontdimen3\font minus
  \fontdimen4\font\relax}
\providecommand{\BIBforeignlanguage}[2]{{%
\expandafter\ifx\csname l@#1\endcsname\relax
\typeout{** WARNING: IEEEtran.bst: No hyphenation pattern has been}%
\typeout{** loaded for the language `#1'. Using the pattern for}%
\typeout{** the default language instead.}%
\else
\language=\csname l@#1\endcsname
\fi
#2}}
\providecommand{\BIBdecl}{\relax}
\BIBdecl

\bibitem{nbody}
M.~S. Warren and J.~K. Salmon, ``A parallel hashed oct-tree ${N}$-body
  algorithm,'' 1993, pp. 12--21.

\bibitem{md}
D.~E. Shaw, R.~O. Dror, J.~K. Salmon, J.~P. Grossman, K.~M. Mackenzie, J.~A.
  Bank, C.~Young, M.~M. Deneroff, B.~Batson, K.~J. Bowers, E.~Chow, M.~P.
  Eastwood, D.~J. Ierardi, J.~L. Klepeis, J.~S. Kuskin, R.~H. Larson,
  K.~Lindorff-Larsen, P.~Maragakis, M.~A. Moraes, S.~Piana, Y.~Shan, and
  B.~Towles, ``Millisecond-scale molecular dynamics simulations on anton,'' in
  \emph{Proceedings of the Conference on High Performance Computing Networking,
  Storage and Analysis}, ser. SC '09.\hskip 1em plus 0.5em minus 0.4em\relax
  New York, NY, USA: Association for Computing Machinery, 2009.

\bibitem{stream}
A.~Arasu, B.~Babcock, S.~Babu, M.~Datar, K.~Ito, I.~Nishizawa, J.~Rosenstein,
  and J.~Widom, ``Stream: the stanford stream data manager (demonstration
  description),'' in \emph{Proceedings of the 2003 ACM SIGMOD International
  Conference on Management of Data}, ser. SIGMOD '03.\hskip 1em plus 0.5em
  minus 0.4em\relax New York, NY, USA: Association for Computing Machinery,
  2003, p. 665.

\bibitem{mpiurl}
M.~Forum, \url{https://www.mpi.org}, 1990.

\bibitem{randomIO}
Y.~Kim, A.~Gupta, B.~Urgaonkar, P.~Berman, and A.~Sivasubramaniam,
  ``Hybridstore: A cost-efficient, high-performance storage system combining
  ssds and hdds,'' in \emph{2011 IEEE 19th Annual International Symposium on
  Modelling, Analysis, and Simulation of Computer and Telecommunication
  Systems}, 2011, pp. 227--236.

\bibitem{dahlin}
M.~D. Dahlin, R.~Y. Wang, T.~E. Anderson, and D.~A. Patterson, ``Cooperative
  caching: using remote client memory to improve file system performance,'' in
  \emph{Proceedings of the 1st USENIX Conference on Operating Systems Design
  and Implementation}, ser. OSDI '94.\hskip 1em plus 0.5em minus 0.4em\relax
  USA: USENIX Association, 1994, p. 19–es.

\bibitem{gms}
M.~J. Feeley, W.~E. Morgan, F.~H. Pighin, A.~R. Karlin, H.~M. Levy, and C.~A.
  Thekkath, ``Implementing global memory management in a workstation cluster,''
  \emph{Proceedings of the fifteenth ACM symposium on Operating systems
  principles}, 1995.

\bibitem{hpccache}
L.~Ou, X.~He, M.~Kosa, and S.~Scott, ``A unified multiple-level cache for high
  performance storage systems,'' in \emph{13th IEEE International Symposium on
  Modeling, Analysis, and Simulation of Computer and Telecommunication
  Systems}, 2005, pp. 143--150.

\bibitem{staging}
R.~Prabhakar, S.~S. Vazhkudai, Y.~Kim, A.~R. Butt, M.~Li, and M.~Kandemir,
  ``Provisioning a multi-tiered data staging area for extreme-scale machines,''
  in \emph{2011 31st International Conference on Distributed Computing
  Systems}, 2011, pp. 1--12.

\bibitem{nvperfr}
IntelTechnologies, \url{https://spdk.io/doc/performance_reports.html}, 2024.

\bibitem{hdf5}
M.~Folk, A.~Cheng, and K.~Yates, ``Hdf5: A file format and i/o library for high
  performance computing applications,'' in \emph{Proceedings of
  Supercomputing}, vol.~99, 1999, pp. 5--33.

\bibitem{markovts}
D.~Joseph and D.~Grunwald, ``Prefetching using markov predictors,'' in
  \emph{Proceedings of the 24th Annual International Symposium on Computer
  Architecture}, ser. ISCA '97.\hskip 1em plus 0.5em minus 0.4em\relax New
  York, NY, USA: Association for Computing Machinery, 1997, p. 252–263.

\bibitem{storagehier}
R.~Mattson, J.~Gecsei, D.~R. Slutz, and I.~L. Traiger, ``Evaluation techniques
  for storage hierarchies,'' \emph{IBM Systems Journal}, vol.~9, no.~2, pp.
  78--117, 1970.

\bibitem{tiered2}
E.~J. O'Neil, P.~E. O'Neil, and G.~Weikum, ``The lru-k page replacement
  algorithm for database disk buffering,'' \emph{SIGMOD Rec.}, vol.~22, no.~2,
  p. 297–306, jun 1993.

\bibitem{tiered3}
G.~Glass and P.~Cao, ``Adaptive page replacement based on memory reference
  behavior,'' in \emph{Proceedings of the 1997 ACM SIGMETRICS International
  Conference on Measurement and Modeling of Computer Systems}, ser. SIGMETRICS
  '97.\hskip 1em plus 0.5em minus 0.4em\relax New York, NY, USA: Association
  for Computing Machinery, 1997, p. 115–126.

\bibitem{zipf}
V.~Almeida, A.~Bestavros, M.~Crovella, and A.~de~Oliveira, ``Characterizing
  reference locality in the www,'' in \emph{Proceedings of the Fourth
  International Conference on on Parallel and Distributed Information Systems},
  ser. DIS '96.\hskip 1em plus 0.5em minus 0.4em\relax USA: IEEE Computer
  Society, 1996, p. 92–107.

\bibitem{iopatterns}
J.~L. Bez, S.~Byna, and S.~Ibrahim, ``I/o access patterns in hpc applications:
  A 360-degree survey,'' \emph{ACM Comput. Surv.}, vol.~56, no.~2, sep 2023.

\bibitem{dahlinps}
P.~Sarkar and J.~H. Hartman, ``Hint-based cooperative caching,'' \emph{ACM
  Trans. Comput. Syst.}, vol.~18, no.~4, p. 387–419, nov 2000.

\bibitem{sumcache}
L.~Fan, P.~Cao, J.~Almeida, and A.~Z. Broder, ``Summary cache: a scalable
  wide-area web cache sharing protocol,'' \emph{SIGCOMM Comput. Commun. Rev.},
  vol.~28, no.~4, p. 254–265, oct 1998.

\bibitem{leff}
A.~Leff, J.~Wolf, and P.~Yu, ``Replication algorithms in a remote caching
  architecture,'' \emph{IEEE Transactions on Parallel and Distributed Systems},
  vol.~4, no.~11, pp. 1185--1204, 1993.

\bibitem{d3n}
E.~U. Kaynar, M.~Abdi, M.~H. Hajkazemi, A.~Turk, R.~Sambasivan, D.~Cohen,
  L.~Rudolph, P.~Desnoyers, and O.~Krieger, ``D3n: A multi-layer cache for the
  rest of us,'' 12 2019, pp. 327--338.

\bibitem{orthus}
K.~Wu, Z.~Guo, G.~Hu, K.~Tu, R.~Alagappan, R.~Sen, K.~Park, A.~C.
  Arpaci-Dusseau, and R.~H. Arpaci-Dusseau, ``The storage hierarchy is not a
  hierarchy: Optimizing caching on modern storage devices with orthus.'' in
  \emph{FAST}, M.~K. Aguilera and G.~Yadgar, Eds.\hskip 1em plus 0.5em minus
  0.4em\relax USENIX Association, 2021, pp. 307--323.

\bibitem{reflex}
A.~Klimovic, H.~Litz, and C.~Kozyrakis, ``Reflex: Remote flash=local flash,''
  in \emph{Proceedings of the Twenty-Second International Conference on
  Architectural Support for Programming Languages and Operating Systems}, ser.
  ASPLOS '17.\hskip 1em plus 0.5em minus 0.4em\relax New York, NY, USA:
  Association for Computing Machinery, 2017, p. 345–359.

\bibitem{Hermes}
A.~Kougkas, H.~Devarajan, and X.-H. Sun, ``I/o acceleration via multi-tiered
  data buffering and prefetching,'' \emph{Journal of Computer Science and
  Technology}, vol.~35, no.~1, pp. 92--120, 2020.

\bibitem{PDC}
J.~Mu, J.~Soumagne, H.~Tang, S.~Byna, Q.~Koziol, and R.~Warren, ``A transparent
  server-managed object storage system for hpc,'' in \emph{2018 IEEE
  International Conference on Cluster Computing (CLUSTER)}, 2018, pp. 477--481.

\bibitem{daos}
M.~Hennecke, ``Understanding daos storage performance scalability,'' in
  \emph{Proceedings of the HPC Asia 2023 Workshops}, ser. HPCAsia '23
  Workshops.\hskip 1em plus 0.5em minus 0.4em\relax New York, NY, USA:
  Association for Computing Machinery, 2023, p. 1–14.

\bibitem{henessay}
J.~L. Hennessy and D.~A. Patterson, \emph{Computer Architecture, Fifth Edition:
  A Quantitative Approach}, 5th~ed.\hskip 1em plus 0.5em minus 0.4em\relax San
  Francisco, CA, USA: Morgan Kaufmann Publishers Inc., 2011.

\bibitem{belady}
A.~V. Aho, P.~J. Denning, and J.~D. Ullman, ``Principles of optimal page
  replacement,'' \emph{J. ACM}, vol.~18, no.~1, p. 80–93, jan 1971.

\bibitem{tarjan}
D.~D. Sleator and R.~E. Tarjan, ``Amortized efficiency of list update and
  paging rules,'' \emph{Commun. ACM}, vol.~28, no.~2, p. 202–208, Feb. 1985.

\bibitem{pagerank}
L.~Page, S.~Brin, R.~Motwani, and T.~Winograd, ``{The PageRank Citation
  Ranking: Bringing Order to the Web},'' Stanford Digital Library Technologies
  Project, Tech. Rep., 1998.

\bibitem{pagemig}
T.~Heo, Y.~Wang, W.~Cui, J.~Huh, and L.~Zhang, ``Adaptive page migration policy
  with huge pages in tiered memory systems,'' \emph{IEEE Transactions on
  Computers}, vol.~71, no.~1, pp. 53--68, 2022.

\bibitem{page1}
G.~S. Paschos, A.~Destounis, L.~Vigneri, and G.~Iosifidis, ``Learning to cache
  with no regrets,'' in \emph{IEEE INFOCOM 2019 - IEEE Conference on Computer
  Communications}, 2019, pp. 235--243.

\bibitem{acme}
I.~Ari, A.~Amer, R.~Gramacy, E.~Miller, S.~Brandt, and D.~Long, ``Acme:
  Adaptive caching using multiple experts,'' \emph{Proceedings in Informatics},
  01 2002.

\bibitem{lruk}
E.~J. O'Neil, P.~E. O'Neil, and G.~Weikum, ``The lru-k page replacement
  algorithm for database disk buffering,'' \emph{SIGMOD Rec.}, vol.~22, no.~2,
  p. 297–306, jun 1993.

\bibitem{WSP}
P.~J. Denning, ``The working set model for program behavior,'' \emph{Commun.
  ACM}, vol.~11, no.~5, p. 323–333, may 1968.

\bibitem{pagelru}
A.~V. Aho, P.~J. Denning, and J.~D. Ullman, ``Principles of optimal page
  replacement,'' \emph{J. ACM}, vol.~18, no.~1, p. 80–93, jan 1971.

\bibitem{score}
G.~Hasslinger, K.~Ntougias, F.~Hasslinger, and O.~Hohlfeld, ``Performance
  evaluation for new web caching strategies combining lru with score based
  object selection,'' in \emph{2016 28th International Teletraffic Congress
  (ITC 28)}, vol.~01, 2016, pp. 322--330.

\bibitem{mlcache}
G.~Vietri, L.~V. Rodriguez, W.~A. Martinez, S.~Lyons, J.~Liu, R.~Rangaswami,
  M.~Zhao, and G.~Narasimhan, ``Driving cache replacement with ml-based
  lecar,'' in \emph{Proceedings of the 10th USENIX Conference on Hot Topics in
  Storage and File Systems}, ser. HotStorage'18.\hskip 1em plus 0.5em minus
  0.4em\relax USA: USENIX Association, 2018, p.~3.

\bibitem{mlcache1}
G.~S. Paschos, A.~Destounis, L.~Vigneri, and G.~Iosifidis, ``Learning to cache
  with no regrets,'' in \emph{IEEE INFOCOM 2019 - IEEE Conference on Computer
  Communications}.\hskip 1em plus 0.5em minus 0.4em\relax IEEE Press, 2019, p.
  235–243.

\bibitem{markovp}
W.~Zucchini and M.~I.L., \emph{Hidden Markov Models for Time Series: An
  Introduction Using R}, 1st~ed.\hskip 1em plus 0.5em minus 0.4em\relax Chapman
  and Hall/CRC, 2009.

\bibitem{mlpatterns}
\BIBentryALTinterwordspacing
N.~Dryden, R.~B\"{o}hringer, T.~Ben-Nun, and T.~Hoefler, ``Clairvoyant
  prefetching for distributed machine learning i/o,'' in \emph{Proceedings of
  the International Conference for High Performance Computing, Networking,
  Storage and Analysis}, ser. SC '21.\hskip 1em plus 0.5em minus 0.4em\relax
  New York, NY, USA: Association for Computing Machinery, 2021. [Online].
  Available: \url{https://doi.org/10.1145/3458817.3476181}
\BIBentrySTDinterwordspacing

\bibitem{queuingsystem}
N.~T. Thomopoulos, \emph{Fundamentals of Queuing Systems}, 2012th~ed.\hskip 1em
  plus 0.5em minus 0.4em\relax Springer New York, 2012.

\bibitem{mlhdd1}
S.~Madireddy, P.~Balaprakash, P.~Carns, R.~Latham, R.~Ross, S.~Snyder, and
  S.~Wild, ``Modeling i/o performance variability using conditional variational
  autoencoders,'' in \emph{2018 IEEE International Conference on Cluster
  Computing (CLUSTER)}, 2018, pp. 109--113.

\bibitem{mllb2}
S.~Madireddy \emph{et~al.}, \emph{Machine Learning Based Parallel I/O
  Predictive Modeling: A Case Study on Lustre File Systems}, 05 2018, pp.
  184--204.

\bibitem{autotuning}
B.~Behzad, J.~Huchette, H.~V.~T. Luu, R.~Aydt, S.~Byna, Y.~Yao, Q.~Koziol, and
  Prabhat, ``A framework for auto-tuning hdf5 applications,'' in
  \emph{Proceedings of the 22nd International Symposium on High-Performance
  Parallel and Distributed Computing}, ser. HPDC '13.\hskip 1em plus 0.5em
  minus 0.4em\relax New York, NY, USA: Association for Computing Machinery,
  2018, p. 127–128.

\bibitem{nvmepf}
L.~Ma, Z.~Liu, J.~Xiong, Y.~Wu, R.~Chen, X.~Peng, Y.~Zhang, G.~Zhang, and
  D.~Jiang, ``zqos: Unleashing full performance capabilities of nvme ssds while
  enforcing slos in distributed storage systems,'' in \emph{Proceedings of the
  53rd International Conference on Parallel Processing}, ser. ICPP '24.\hskip
  1em plus 0.5em minus 0.4em\relax New York, NY, USA: Association for Computing
  Machinery, 2024, p. 618–628.

\bibitem{libra}
\BIBentryALTinterwordspacing
D.~Shue and M.~J. Freedman, ``From application requests to virtual iops:
  provisioned key-value storage with libra,'' in \emph{Proceedings of the Ninth
  European Conference on Computer Systems}, ser. EuroSys '14.\hskip 1em plus
  0.5em minus 0.4em\relax New York, NY, USA: Association for Computing
  Machinery, 2014. [Online]. Available:
  \url{https://doi.org/10.1145/2592798.2592823}
\BIBentrySTDinterwordspacing

\bibitem{cachemapping}
\BIBentryALTinterwordspacing
G.~H. Golub and C.~F. Van~Loan, \emph{Matrix Computations - 4th Edition}.\hskip
  1em plus 0.5em minus 0.4em\relax Philadelphia, PA: Johns Hopkins University
  Press, 2013. [Online]. Available:
  \url{https://epubs.siam.org/doi/abs/10.1137/1.9781421407944}
\BIBentrySTDinterwordspacing

\bibitem{mercury}
J.~Soumagne, D.~Kimpe, J.~Zounmevo, M.~Chaarawi, Q.~Koziol, A.~Afsahi, and
  R.~Ross, ``Mercury: Enabling remote procedure call for high-performance
  computing,'' in \emph{2013 IEEE International Conference on Cluster Computing
  (CLUSTER)}, 2013, pp. 1--8.

\bibitem{onlinelearning}
S.~Shalev-Shwartz, ``Online learning and online convex optimization,''
  \emph{Found. Trends Mach. Learn.}, vol.~4, no.~2, p. 107–194, feb 2012.

\bibitem{talus}
N.~Beckmann and D.~Sanchez, ``Talus: A simple way to remove cliffs in cache
  performance,'' in \emph{2015 IEEE 21st International Symposium on High
  Performance Computer Architecture (HPCA)}, 2015, pp. 64--75.

\bibitem{fileprefetching}
\BIBentryALTinterwordspacing
A.~J. Smith, ``Sequentiality and prefetching in database systems,'' \emph{ACM
  Trans. Database Syst.}, vol.~3, no.~3, p. 223–247, Sep. 1978. [Online].
  Available: \url{https://doi.org/10.1145/320263.320276}
\BIBentrySTDinterwordspacing

\bibitem{delta}
NCSA, \url{https://docs.ncsa.illinois.edu/systems/delta/en/latest/}, 2023.

\bibitem{ws}
A.~Blum and C.~Burch, ``On-line learning and the metrical task system
  problem,'' in \emph{Proceedings of the Tenth Annual Conference on
  Computational Learning Theory}, ser. COLT '97.\hskip 1em plus 0.5em minus
  0.4em\relax New York, NY, USA: Association for Computing Machinery, 1997, p.
  45–53.

\bibitem{costmodel}
G.~Yadgar, M.~Factor, and A.~Schuster, ``Cooperative caching with return on
  investment,'' in \emph{2013 IEEE 29th Symposium on Mass Storage Systems and
  Technologies (MSST)}, 2013, pp. 1--13.

\bibitem{kelly}
F.~P. Kelly, \emph{Reversibility and Stochastic Networks}.\hskip 1em plus 0.5em
  minus 0.4em\relax USA: Cambridge University Press, 2011.

\bibitem{queuingtheory}
G.~Grimmett and D.~Stirzaker, \emph{Probability and random processes}.\hskip
  1em plus 0.5em minus 0.4em\relax Oxford; New York: Oxford University Press,
  2001.

\bibitem{logp}
D.~Culler, R.~Karp, D.~Patterson, A.~Sahay, K.~E. Schauser, E.~Santos,
  R.~Subramonian, and T.~von Eicken, ``Logp: towards a realistic model of
  parallel computation,'' \emph{SIGPLAN Not.}, vol.~28, no.~7, p. 1–12, Jul.
  1993.

\bibitem{statistics}
T.~Hastie, R.~Tibshirani, and J.~Friedman, \emph{The elements of statistical
  learning: data mining, inference and prediction}, 2nd~ed.\hskip 1em plus
  0.5em minus 0.4em\relax Springer, 2009.

\bibitem{Rmanual}
\BIBentryALTinterwordspacing
{R Core Team}, \emph{R: A Language and Environment for Statistical Computing},
  R Foundation for Statistical Computing, Vienna, Austria, 2021. [Online].
  Available: \url{https://www.R-project.org/}
\BIBentrySTDinterwordspacing

\bibitem{libaio}
Linux, \url{https://pagure.io/libaio}, 2017.

\bibitem{spdk}
Intel, \url{https://spdk.io/doc/}, 2024.

\bibitem{perfmodelgpu}
A.~Sasidharan, ``Performance models for hybrid programs accelerated by gpus,''
  in \emph{2021 IEEE International Parallel and Distributed Processing
  Symposium Workshops (IPDPSW)}, 2021, pp. 641--651.

\bibitem{gc}
X.-Y. Hu, E.~Eleftheriou, R.~Haas, I.~Iliadis, and R.~Pletka, ``Write
  amplification analysis in flash-based solid state drives,'' in
  \emph{Proceedings of SYSTOR 2009: The Israeli Experimental Systems
  Conference}, ser. SYSTOR '09.\hskip 1em plus 0.5em minus 0.4em\relax New
  York, NY, USA: Association for Computing Machinery, 2009.

\bibitem{mlhdd2}
M.~Wang, K.~Au, A.~Ailamaki, A.~Brockwell, C.~Faloutsos, and G.~Ganger,
  ``Storage device performance prediction with cart models,'' in \emph{The IEEE
  Computer Society's 12th Annual International Symposium on Modeling, Analysis,
  and Simulation of Computer and Telecommunications Systems, 2004. (MASCOTS
  2004). Proceedings.}, 2004, pp. 588--595.

\bibitem{mlhdd3}
J.~Kunkel, M.~Zimmer, and E.~Betke, ``Predicting performance of non-contiguous
  i/o with machine learning,'' 07 2015, pp. 257--273.

\bibitem{slingshot}
K.~Shafie~Khorassani, C.~C. Chen, B.~Ramesh, A.~Shafi, H.~Subramoni, and
  D.~Panda, ``High performance mpi over the slingshot interconnect: Early
  experiences,'' in \emph{Practice and Experience in Advanced Research
  Computing 2022: Revolutionary: Computing, Connections, You}, ser. PEARC
  '22.\hskip 1em plus 0.5em minus 0.4em\relax New York, NY, USA: Association
  for Computing Machinery, 2022.

\bibitem{tbb}
C.~Pheatt, ``Intel® threading building blocks,'' \emph{J. Comput. Sci. Coll.},
  vol.~23, no.~4, p. 298, apr 2008.

\bibitem{pmdk}
Intel,
  \url{https://www.intel.com/content/www/us/en/developer/articles/technical/introducing-the-persistent-memory-development-kit.html},
  2020.

\bibitem{lfumodel}
S.~Vanichpun and A.~M. Makowski, ``The output of a cache under the independent
  reference model: where did the locality of reference go?'' \emph{SIGMETRICS
  Perform. Eval. Rev.}, vol.~32, no.~1, p. 295–306, jun 2004.

\end{thebibliography}

\end{document}